\documentclass{elsart}

\journal{Chemical Physics Letters}
\usepackage{graphics}
\usepackage{epsfig}

\usepackage{amssymb}

\def\BNF{B$_{24}$N$_{24}$ }
\def\Sf{{\sl S$_4$} }
\def\O{{\sl O} }
\def\Se{{\sl S$_8$} }
\def\BNE{B$_{28}$N$_{28}$ }
\def\BNT{B$_{32}$N$_{32}$ }
\def\ocm{cm$^{-1}$ }

\begin{document}

\begin{frontmatter}



\title{ Electronic structure, vibrational stability,  infra-red, and Raman spectra of \BNF cages}
\author[GW]{Rajendra R. Zope \corauthref{ZCOR}}
\ead{rzope@alchemy.nrl.navy.mil}
\author[GT,MNRL]{Tunna Baruah }
\ead{baruah@dave.nrl.navy.mil}
\author[MNRL]{Mark R. Pederson}
\ead{pederson@dave.nrl.navy.mil}
\author[NRL]{Brett  I. Dunlap}
\ead{dunlap@nrl.navy.mil}

\corauth[ZCOR]{Fax: +1-202-767-1716}

\address[GW]{Department of Chemistry, George Washington University, Washington DC, 20052, USA}

\address[GT]{Department of Physics, Georgetown University, Washington DC, 20057, USA}

\address[MNRL]{Center for Computational Materials Science, US Naval Research Laboratory, Washington DC 20375, USA}

\address[NRL]{Code 6189, Theoretical Chemistry Section, US Naval Research Laboratory, Washington, DC 20375, USA}
\date{\today}

\begin{abstract}
   We examine the vibrational stability of three candidate structures for the \BNF cage
and report their infra-red  (IR) and Raman spectra. The candidate structures considered 
are a round cage with octahedral \O symmetry, a cage with \Sf symmetry that satisfies 
the isolated square rule, 
and a cage of \Se symmetry, which combines the caps of the (4,4) nanotube, 
and contains two extra squares and octagons.
The calculations are performed within density functional theory, at the all electron level, with large basis sets,
and within the generalized gradient approximation. The vertical ionization potential (VIP) 
and static dipole polarizability are also reported.  The \Sf and \Se cages are
energetically nearly degenerate and are favored over the \O cage which has 
six extra octagons and squares.  The IR and Raman spectra of the three 
clusters show notable differences providing thereby a way to identify 
and possibly synthesize the cages.

\end{abstract}


\end{frontmatter}

\begin{figure}
\epsfig{file=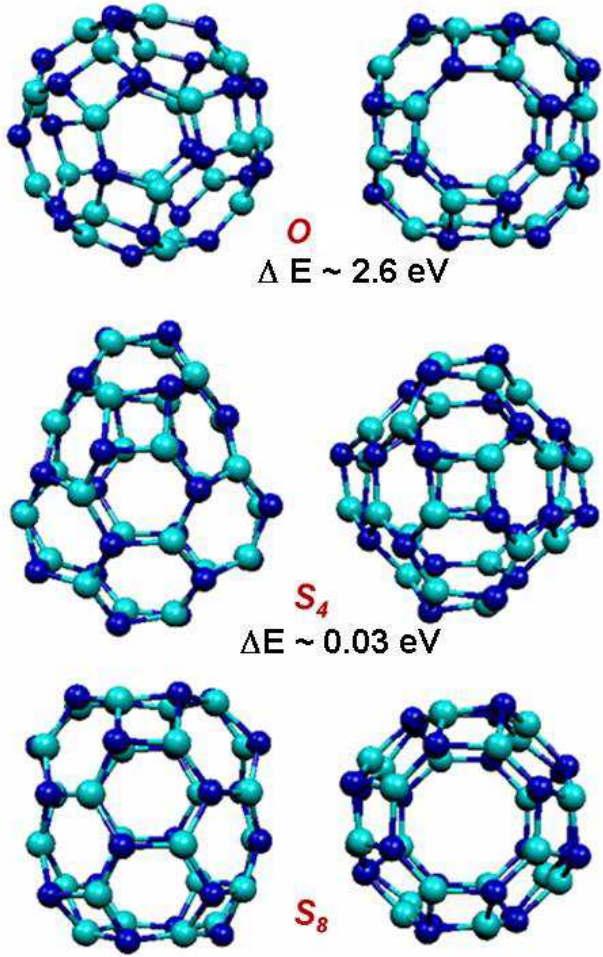,width=8.5cm,clip=true}
\caption{\label{fig1} (Color online) Two different views of  optimized 
\BNF cages.
The relative energies are given with respect to S$_8$ total energy.}
\end{figure}

   Recently, boron nitride  (BN) cages were synthesized and detected by laser desorption time of flight
mass spectroscopy \cite{Oku03}. The \BNF cluster was observed in abundance and its structure was proposed to 
be a round cage structure with octahedral (\O) symmetry (Cf. Fig. ~\ref{fig1}), in an analogy to 
the icosahedral C$_{60}$ cluster.  It consists of alternate BN atoms and is made round by introducing
eighteen defects: six octagons and twelve squares, in the hexagonal network of a single BN sheet.
In the icosahedral C$_{60}$ fullerene every carbon atom is equivalent by symmetry.
Similarly  every boron or nitrogen atom in this octahedral structure is equivalent by symmetry.
Taking a cue from the similarity of C$_{60}$ fullerene and the O \BNF cage it was suggested that 
the two halves of the round octahedral cage can form hemispherical caps for the (4,4) BN
nanotubes \cite{ZB04} analogous to C$_{60}$ hemispheres capping the (5,5) carbon nanotube \cite{MDW92}.  
Further by Euler theorem, in analogy to the twelve isolated pentagons of C$_{60}$,  six isolates squares  
close the alternate BN fullerenes \cite{ZSK97,SSL95}.  For the B$_{24}$N$_{24}$, the resultant 
cage  (Cf. Fig. ~\ref{fig1})
has \Sf symmetry \cite{SSL95}. 
In a  recent  study \cite{WJ04} that examined seven candidate structures for the \BNF geometry, 
the \Sf  cage  was found to be energetically favorable over the \O cage in agreement with our work.
A new, most stable cage structure was found that  has \Se symmetry and 
contains eight squares and two octagons in the hexagonal network \cite{WJ04}. 
The \Se cage is related to the \BNE or \BNT  (4,4) closed nanotubes in Ref. \cite{ZB04}
and can be derived
from \BNE by eliminating  a middle  ring of alternating BN atoms followed by rotation 
by 45$^o$.  The energy differences
between these isomers, particularly between the \Se and \Sf cages are quite small (0.1eV or less)
and thus based on energetics 
it is not apparent which isomer is most likely to be observed in the experiment. 
Clearly  more theoretical calculations or further experimental characterization is necessary.
 For this purpose we have carried out density functional \cite{KS65}  calculations on the three
candidate structures of the \BNF clusters. We predict the infra-red (IR) and Raman spectra for the
three isomers  which may aid the experimentalists in identifying the structure
of \BNF. Along with these spectra we also report the ionization potential, 
and the static dipole polarizability of the three cages. Our calculations are carried out
at the all electron level by the { NRLMOL} code \cite{NRLMOL1,NRLMOL2,NRLMOL3}, using large polarized Gaussian 
basis sets \cite{Porezag99} optimized for these density functional calculations, and 
using the Perdew-Burke-Ernzerhof   generalized gradient approximation (PBE-GGA) \cite{PBE}. 

\begin{table}
\caption{The total electronic energy E (in Hartree), the zero point energy (ZPE), $\Delta \,E$ (energy relative 
to the energy of \Se cage and excluding ZPE contribution)  (in kJ/mol), binding energy (BE) per atom (in eV), the vertical ionization potential (VIP) (in eV), and the HOMO-LUMO gap (in eV)
of the \BNF cages. The ZPE is half the sum of vibrational frequencies.}
\begin{tabular}{llccccc}
\hline
             & E       & ZPE  &  $\Delta E$    & BE    & VIP  & gap  \\
\hline         
\O         & -1910.96994  &  664.49  &   2.6   &  8.27 &  8.4   &  4.7     \\
\Sf        & -1911.06552  &  678.74  &   0.03  &  8.32 &  8.5   &  4.9     \\
\Se        & -1911.06676  &  675.02  &   0.0   &  8.32 &  8.3   &  4.6       \\
\hline
\end{tabular}
\label{tab:energetics}
\end{table}

         The energetics and the electronic properties of optimized structures are 
summarized in Table \ref{tab:energetics}.  The calculations predict the \Se cage as the 
lowest energy structure. The \Sf isomer is within 0.03 eV while the \O cage is 2.6 eV
higher. This energy ordering is in accord with earlier hybrid density functional (B3LYP/6-31G*)
calculations \cite{WJ04}. These energy differences are of same  magnitude 
even with addition of the  zero-point energy.
The vibrational frequencies determined within the harmonic approximation indicate all structures
to be local minima on the potential energy surface.  Further, consistent with earlier 
prediction on the BN cages and nanotubes \cite{XVCC98}, these cages are characterized by a wide energy 
gap of about 4.7 eV between the highest occupied molecular orbital (HOMO) and the lowest unoccupied 
molecular orbital (LUMO).  We note that the HOMO-LUMO gaps obtained within the present PBE-GGA 
generally underestimate the sum of ionization potential and electron affinity.
The vertical ionization potential (VIP) is calculated from the
the differences in total energy of the self-consistent solution of the cluster and 
its singly charged positive ion, at the optimal ionic configuration of the neutral 
cluster. The calculated VIP are also given in Table \ref{tab:energetics}.  
We note that the VIP reported for the \O \BNF cage in Ref. \cite{ZB04} should 
be 8.64 eV instead of incorrectly reported value of 6.66 eV.  The  vertical electron affinity 
calculations can be performed in similar fashion. We find that the 
additional electron is weakly bound (despite relatively large basis set used in the 
calculation its eigenvalue turns out to be positive
although the total energy is lowered with respect to neutral cluster).  

\begin{table}
\caption{The static dipole polarizability (in {\AA}$^3$)  of the \O, \Sf, and \Se isomers of \BNF.}
\begin{tabular}{llcll}
\hline
             & $\alpha_{xx}$   & $\alpha_{yy}$ & $\alpha_{zz}$  & ${\bar {\alpha}}$ \\
\hline         
\O         &   53.8  & 53.8  &  53.8  &  53.8  \\
\Sf        &   49.3  &  49.3  & 55.6  &  51.4  \\
\Se        &   50.7  & 50.7  &  55.4  & 52.3  \\
\hline
\end{tabular}
\label{tab:pol}
\end{table}
  The static dipole polarizability is an important physical quantity that characterizes
the electric response of the clusters to applied uniform electric field. It is calculated
from the so called {\sl finite field} method in which the self-consistent problem 
is solved for the usual Hamiltonian augmented with the field term $\vec{E}\cdot\vec{r}$.
The polarizability tensor is built from the self-consistent solutions performed for
various values of electric field ${E}$ along different directions. The nuclei 
are assumed to be frozen.  The calculated values are shown in Table \ref{tab:pol}. 
The mean polarizability of the three isomers is of 
similar magnitude with the \O cage being slightly more polarizable. The structural 
differences of the isomers are reflected in the anisotropy of the polarizability.
Unlike the perfectly round \O cage, the isomers \Se and \Sf have nonzero anisotropy. 

\begin{table}
\caption{The IR active frequencies of \BNF cages. The quantity inside the bracket is the intensity (Debye$^2$/amu/\AA$^2$)
the respective absorption. Only IR active modes with intensity greater than 3 (Debye$^2$/amu/\AA$^2$) are given.}
\begin{tabular}{lccc}
\hline
             & Symmetry      & Frequency (\ocm)  \\
\hline         
\O         &   T$_1$   &  759 (7), 1471 (83) \\ 

\Sf        &   B       &  1368 (7),  1423 (47), 1448 (34) \\
           &   E       &  1353 (7),  1399 (26),  1442 (30) \\

\Se        &   B       &  1432 (87) \\
           &   E$_1$   &  767 (7), 1406 (27), 1461 (45).  \\
\hline
\end{tabular}
\label{tab:IR}
\end{table}

\begin{figure}
\epsfig{file=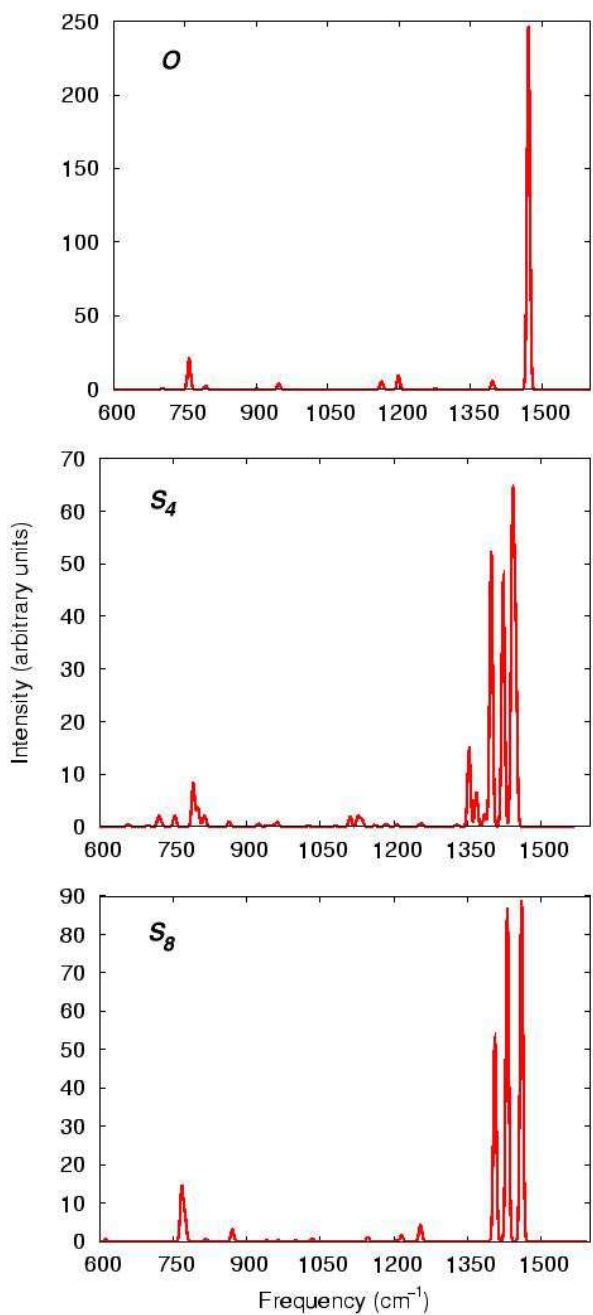,width=8.4cm,clip=true}
\caption{\label{fig:IR} The infra-red spectra of the three cage structures  of  the \BNF.}
\end{figure}

      The calculated IR  spectra are presented in Fig. ~\ref{fig:IR}.  The frequencies
and their symmetries of IR active modes are listed in Table \ref{tab:IR}. 
The IR absorption intensities are broadened by 6 \ocm to mimic experimental uncertainties.
The IR spectra of the three isomers have  different structure.
The high symmetry of the round \O cage results in its IR spectrum with four peaks.
It shows a strong absorption at 1471 \ocm and a weak absorption at 759 \ocm. 
Other peaks around 950 \ocm and 500 \ocm have relative intensity less than 2\%. 
The strong absorption mode at 1471 \ocm is three fold degenerate and corresponds to 
stretching and a compression of the alternate bond in the octagon.
An earlier IR calculation \cite{P2000} on the \O  \BNF cage that used semi-empirical modified neglect of diatomic
overlap (MNDO) method reports three very different IR active modes  at 1356, 1336, and 772 \ocm.
The IR spectra of \Sf can be broadly classified into two categories of bands in the frequency
range 1353-1448 \ocm.  The first set consists of small peaks around 1350 \ocm
while the second set shows three peaks at 1399, 1423, and at 1442-1448 \ocm.  In the
experiment, the later set of peaks may show up as a broad band beginning with a shoulder around 1350 \ocm.
The IR spectra of the \Se structure exhibits three conspicuous peaks at  1407, 1432, and 1461 \ocm.
The strongest mode at 1432 \ocm corresponds to a compression of two opposite hexagonal bonds accompanied
by simultaneous stretching of remaining hexagonal bonds. The absorption at 1461 \ocm 
is due to the  compression and stretching of bonds of octagons. 
A small peak is also observed at lower frequency of 767 \ocm. It corresponds to a vibrational mode
that results in puckering of alternate atoms. 
The differences observed in the structure of the predicted IR spectra of the three isomers indicate the
possibility of identifying the isomer by high resolution IR spectroscopy measurement. A sharp, high-frequency
peak would indicate the presence of the octahedral isomer.


\begin{figure}
\epsfig{file=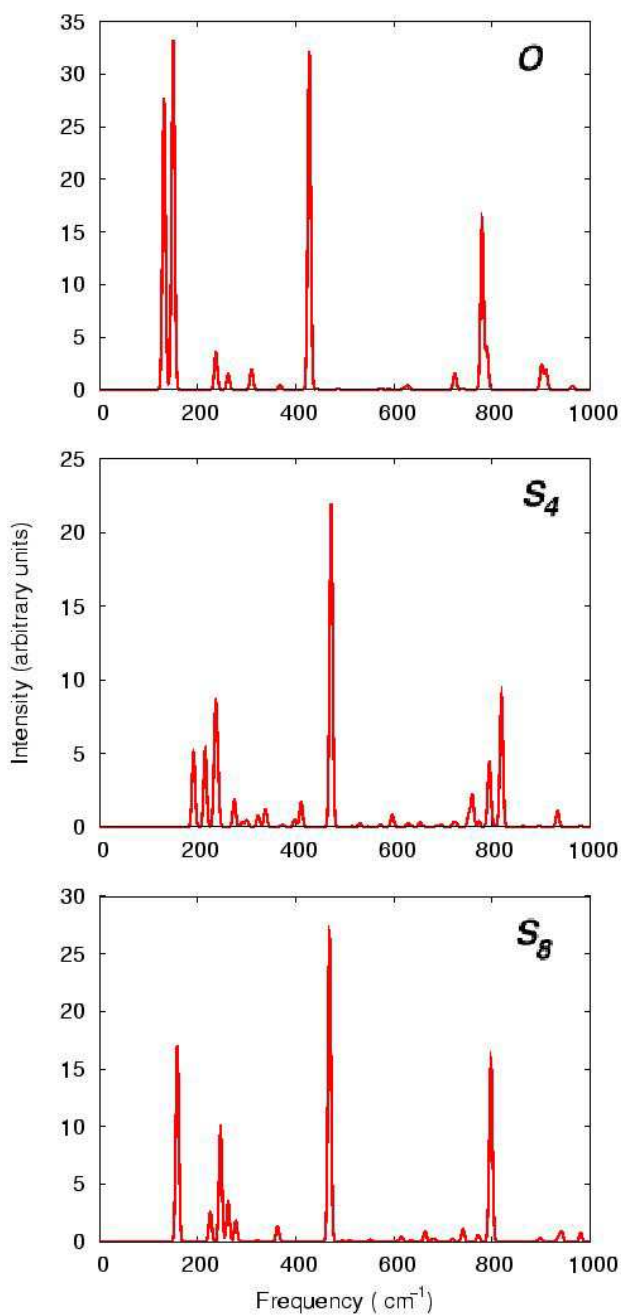,width=8.4cm,clip=true}
\caption{\label{fig:Raman} The Raman spectra of the three cage structures  of  the \BNF.}
\end{figure}

\begin{table}
\caption{The Raman active frequencies of \BNF cages. 
(Frequencies with very weak absorption are not presented).}
\begin{tabular}{lccc}
\hline
             & Symmetry      & Frequency (\ocm)  \\
\hline         
\O         &   T$_2$   &    150 \\ 
           &   A$_1$   &   428, 780, 790, 901 \\
           &   E       &   131, 910  \\
\Sf        &   A       &   192, 215, 472, 760, 794, 820 \\
           &   E       &    236  \\
\Se        &   A       &   262, 469, 798 \\
           &   E$_2$   &   158, 247 \\
\hline
\end{tabular}
\label{tab:Raman}
\end{table}

 The Raman spectra of the three \BNF cages are shown in Fig. ~\ref{fig:Raman} 
and the vibrational frequencies at which significant absorption occurs are presented
in Table \ref{tab:Raman}.
The Raman spectra of three cages show an interesting feature:
all three cages show an intense absorption in the 420 - 480 \ocm region.
In all the three structures this mode is an A mode and corresponds to
the breathing motion of the cages. The intensity of this peak is comparable 
in all the three structures.
The \O cage is  marked by an intense double peak in the 131-150 \ocm 
region which is not seen in the other two spectra. These peaks are associated
with $E$ and  $T_2$ modes.
Thus two peaks below 160 \ocm and another one  at around 430 \ocm, all with 
similar intensity will signify a \O structure.  
On the other hand, the \Sf structure will show  three low Raman peaks in
the region 190 - 230 \ocm and three others around 800 \ocm.
In experiment, they may appear as broad peaks with roughly half 
the intensity of the most intense peak at 430 \ocm. The \O structure 
also shows a weak absorption around 800 \ocm.
The S8 structure has a low frequency peak around 158 \ocm which 
is about half as intense as the peak at 472 \ocm, and a 
similar one at 798 \ocm.  The peak at 158 \ocm due to a doubly
degenerate  E mode while the one at 798 \ocm has A symmetry.  Further it 
will show a broad band  at 230 \ocm which will be missing in the 
two other structures.  These differences in the Raman spectra of the
 \O, \Sf, and \Se cages may help in identifying them in 
the experimental spectra.

       To summarize, three candidate structures for the ground state geometry 
of the \BNF are considered. These are, the octahedral structure proposed by the 
experimentalists, the \Sf cage that satisfy the isolated square rule, and the 
\Se symmetric tubule containing two octahedrons and eight squares. 
The harmonic frequency analysis indicate all structures  to be vibrationally stable.
The \Se tubule and \Sf are energetically nearly 
degenerate and are favored over the \O cage on the basis of energetics.
All the three clusters have wide HOMO-LUMO gap and high ionization potential.
The infra-red and Raman spectra show notable differences and therefore point to possible
identification of the  structure  by the IR or Raman spectroscopy. The IR spectra perhaps,
as in the case of C$_{60}$ \cite{WJG87,KFH90}, could guide methods for optimizing the production of the
round cluster.

        The Office of Naval Research, directly and through the Naval Research Laboratory, and the 
Department of Defense's   High Performance Computing Modernization Program, through the Common High 
Performance Computing Software Support Initiative Project MBD-5, supported this work.


\begin{thebibliography}{00}

\bibitem{Oku03} T. Oku, A. Nishiwaki, I. Narita, M. Gonda, Chem. Phys. Lett. {\bf  380} (2003) 620.

\bibitem{ZB04} R. R. Zope,  B. I. Dunlap, Chem. Phys. Lett. {\bf  386} (2004) 403.

\bibitem{MDW92} J. W. Mintmire, B. I. Dunlap, C. T. White, Phys. Rev. Lett.  {\bf 68} (1992) 631.

\bibitem{ZSK97} H.-Y. Zhu, T. G. Schmalz, D. J. Klein, Int. J. Quantum Chem.  {\bf 63} (1997) 393.

\bibitem{SSL95} M.-L. Sun, Z. Slanina, S.-L. Lee, Chem. Phys. Lett.  {\bf 233} (1995) 279.

\bibitem{WJ04} H. Wu,  H. Jiao, Chem. Phys. Lett. {\bf  386} (2004) 369.


\bibitem{P2000} V. V. Pokropivny, V. V. Skorokhod, G. S. Oleinik, A. V. Kurdyumov, T. S. Bartnitskaya,
A. V. Pokropivny, A. G. Sisonyuk, D. M. Sheichenko,  J. Solid State Chem. {\bf 154} (2000) 214.

\bibitem{KS65} W. Kohn, L. J. Sham, Phys. Rev.  {\bf 140} (1965) A1133.

\bibitem{NRLMOL1}
    M. R. Pederson, K. A. Jackson, Phys. Rev. B. {\bf 41}, 7453 (1990).
\bibitem{NRLMOL2}
    K. A. Jackson, M. R. Pederson, Phys. Rev. B. {\bf 42}, 3276 (1990).
\bibitem{NRLMOL3}
    M. R. Pederson, K. A. Jackson, Phys. Rev. B. {\bf 43}, 7312 (1991).

\bibitem{Porezag99}
D. V. Porezag, M. R. Pederson, Phys. Rev. A {\bf 60}, 2840 (1999).

\bibitem{PBE}
  J. P. Perdew, K. Burke, M. Ernzerhof,  Phys.  Rev.  Lett.   {\bf  77 }, 3865(1996).

\bibitem{XVCC98} X. Blase, A. De Vita, J.-C. Charlier, R. Car, Phys. Rev. Lett.  {\bf 80} (1998) 1666.

\bibitem{WJG87} Z. C. Wu, D. A. Jelski, and T. F. George,  Chem. Phys. Lett. {\bf 137}, (1987) 291.

\bibitem{KFH90} W. Kr\"atschmer, K. Fostiropoulous, D. R. Huffman,  Chem. Phys. Lett. {\bf 170}, 
(1990) 167.
\end{thebibliography}
\end{document}